\documentclass[prb,aps,twocolumn,showpacs]{revtex4-1}
\usepackage{graphicx}% Include figure files
\usepackage{dcolumn}% Align table columns on decimal point
\usepackage{bm}% bold math
\usepackage{dsfont}
\usepackage{color}  
\usepackage{dcolumn}
\usepackage{array}
\usepackage{xcolor}
\usepackage{graphicx}
\usepackage{amsmath}
\usepackage{commath}
\usepackage{bbold}
\usepackage{textcomp}
\usepackage{hyperref}
\hypersetup{
 colorlinks=true,
 linkcolor=blue,
 filecolor=magenta,
 urlcolor=cyan}

\begin{document}

\title{Current-driven skyrmion Depinning in Magnetic Granular Films}
\author{A. Salimath$^1$}
\email{akshaykumar.salimath@kaust.edu.sa}
\author{A. Abbout$^1$}
\author{A. Brataas$^2$}
\author{A. Manchon$^1$}
\email{aurelien.manchon@kaust.edu.sa}
\affiliation{$^1$King Abdullah University of Science and Technology (KAUST),
Physical Science and Engineering Division (PSE), Thuwal 23955-6900, Saudi Arabia}
\affiliation{$^2$Center for Quantum Spintronics, Department of Physics, Norwegian University of Science and Technology, NO-7491 Trondheim, Norway}
%\date{\today}

\begin{abstract}
We consider current-driven motion of magnetic skyrmions in granular magnetic films. The study uses micromagnetic modeling and phenomenological analysis based on the Thiele formalism. Remarkably, disorder enhances the effective skyrmion Hall effect that depends on the magnitude of the driving force (current density and non-adiabaticity parameter). The origin is sliding motion of the skyrmion along the grain boundaries, followed by pinning and depinning at the grain junctions. A side-jump can occur during this depinning process. In addition, the critical current that triggers the skyrmion motion depends on the relative size of the crystallites with respect to the skyrmion size. Finally, when the skyrmion trajectory is confined along an edge by the non-adiabatic Magnus force, the critical current density can be significantly reduced. Our results imply that narrow nanowires have higher skyrmion mobilities.
\end{abstract}

\maketitle
\section{Introduction} 

{\color{red}Twisted spin textures called skyrmions \cite{1,2,21}, obtained in chiral magnets with bulk \cite{3,4,5,6,7,8} or interfacial Dzyaloshinskii-Moriya interaction (DMI) \cite{9,14,10,12,Belabbes}, and recently observed at room temperature \cite{Jiang2015,Chen2015,13,15,16}, are promising candidates for the next generation digital storage and processing applications \cite{17,19,Kang2016}. Besides racetrack memory devices \cite{18,Zhu2017}, skyrmions have been proposed as building blocks for transistors \cite{Zhang2015}, logic applications \cite{Luo2018,Zhang,Mankalale2018} and neuromorphic computing \cite{Huang2017,Li2017,Chen2018,Pinna2018}.} Their rich topological properties combined with the ultralow energy required to manipulate them give them an edge over their predecessors such as domain walls and magnetic vortices \cite {17,18, 20,21,22,23,24,25}. Since their discovery, considerable research has been devoted to understand their topological properties, room temperature stability and robustness against disorder \cite{15,16,17,20,21,22,23,24,25,26}. Still, a deep understanding of their dynamic behavior in real thin films characterized by granular boundaries, is lacking. Magnetic/non-magnetic thin films used in technological applications are deposited using magneton sputtering techniques and, therefore, exhibit polycrystallinity, resulting in fluctuations of the magnetic properties such as magnetic anisotropy or DMI \cite {26,27}. In micromagnetic simulations, the magnetic systems are typically assumed homogeneous and the effect of the material parameters variations are disregarded. \par

However, the most recent experiments show that disorder substantially impacts current-driven motion of magnetic solitons (domain walls and skyrmions) and cannot be neglected \cite{26,28,Litzius2017}. While previous studies have focused on the effect of confining potentials \cite {29,Iwasaki2014} and random point defects \cite {31,32,33,34,Woo}, a detailed description of the skyrmion motion in polycrystalline systems remains lacking. In the present work, we investigate current-driven motion of skyrmions in polycrystalline samples using micromagnetic modeling and phenomenological analysis. We find that the disorder enhances the effective skyrmion Hall effect that depends on the magnitude of the driving force (current density and non-adiabaticity parameter). It is the sliding motion of the skyrmion along the grain boundaries that causes this enhancement, as confirmed by simulations and Thiele's rigid motion model. In addition, our simulations show that the critical depinning current depends on the relative size of the crystallite with respect to the skyrmion size. The skyrmion motion is less influenced by the pinning forces when its motion is confined by the sample edge.

\section{skyrmion motion in polycrystalline materials}
\subsection{Micromagnetic simulation setup}
The micromagnetic simulations are performed using the OOMMF open source software package \cite{OOMMF}. In the present work, the motion is induced by spin transfer torque, as described by the Landau-Lifshitz-Gilbert equation \cite{STT}, 
\begin{multline}
\partial_t\mathbf{m}=-\gamma\mu_0{\bf m}\times\mathbf{H}_{\rm eff}+ \alpha\mathbf{m}\times\partial_t\mathbf{m}-v_{s}\partial_x\mathbf{m} \\ + \beta v_{s}\mathbf{m} \times \partial_x\mathbf{m},\end{multline}%
where $\mu_0\mathbf{H}_{\rm eff}=-(1/M_s)\delta W/\delta {\bf m}$ is the effective field, $W$ being the magnetic energy density and $M_s$ the saturation magnetization, $v_{s}=(J_ePg\mu_{B})/(2eM_s)$ is the driving electron velocity, $J_e$ is the current density applied along ${\bf x}$, $P$ is the spin polarization and $\beta$ is the non-adiabicity parameter. Electrons flow towards the right for $v_{s}>$ 0. The energy density $W$ is

\begin{multline}\label{eq2}
W=A\sum_i(\partial_i{\mathbf m})^2-K(\mathbf n\cdot \mathbf m)^{2}- \frac{\mu_{0}}{2}M_{s} \mathbf m\cdot\mathbf H_{d}-\mu_{0}M_{s} \mathbf m\cdot\mathbf H \\ - D_{\rm DM}{\bf m}\cdot[({\bf z}\times{\bm\nabla})\times{\bf m}].\end{multline}
Here $A$ is the magnetic exchange, $K$ is the uniaxial magnetic anisotropy constant along direction ${\bf n}$, $\bf H$ is the applied magnetic field and ${\bf H}_{d} $ is the demagnetization field, while $D_{\rm DM}$ is the DMI parameter. \par

In the simulations, we consider a ferromagnetic layer of size 400$\times$400$\times$1 nm$^3$ with a perpendicular uniaxial anisotropy. The layer is discretized into cells of volume 2$\times$2$\times$1/3 nm$^3$. The material's parameters are $M_s=580$ kA/m, $A$=15 pJ/m, $K=0.6$ MJ/m$^{3}$, and $D_{\rm DM}=3.0$ mJ/m$^{2}$. {\color{red} These parameters correspond to the experimental values for a Co/Pt system \cite{25}.} An external magnetic field $\mu_0H$= 125 mT is applied perpendicular to the sample plane to stabilize the skyrmion which is relaxed at a distance of 100 nm from the left edge of the sample. For these parameters, a relaxed skyrmion size of 18 nm is obtained. A large damping parameter of $\alpha$=0.5 is considered in the study, close to the experimental conditions \cite{Litzius2017}.  {\color{red}Considering a lower value of damping coefficient does not alter our final conclusions, but only results in a shift in critical current (not shown).} The disorder is modeled as crystallites with uniform magnetic anisotropy strength and varying anisotropy direction. {\color{red}These crystallites are generated following a random grain map obtained from Voronoi tessellation \cite{Lau,Touko}.} The anisotropy axis assigned to each grain is picked randomly within the conical region symmetric about the ${\bf z}$-axis with a semi angle defined by $\theta$. $\theta$ defines the disorder strength. All micromagnetic simulations have been averaged over 10 to 25 realizations for each configuration of grain size $\lambda$ and disorder strength $\theta$. {\color{red} The effects of temperature are left for future work (see section \ref{s:thcom}).}\par

\subsection{Enhanced skyrmion Hall effect and sliding motion}
\begin{figure}[ht]
\begin{center}
\includegraphics[width=0.48\textwidth]{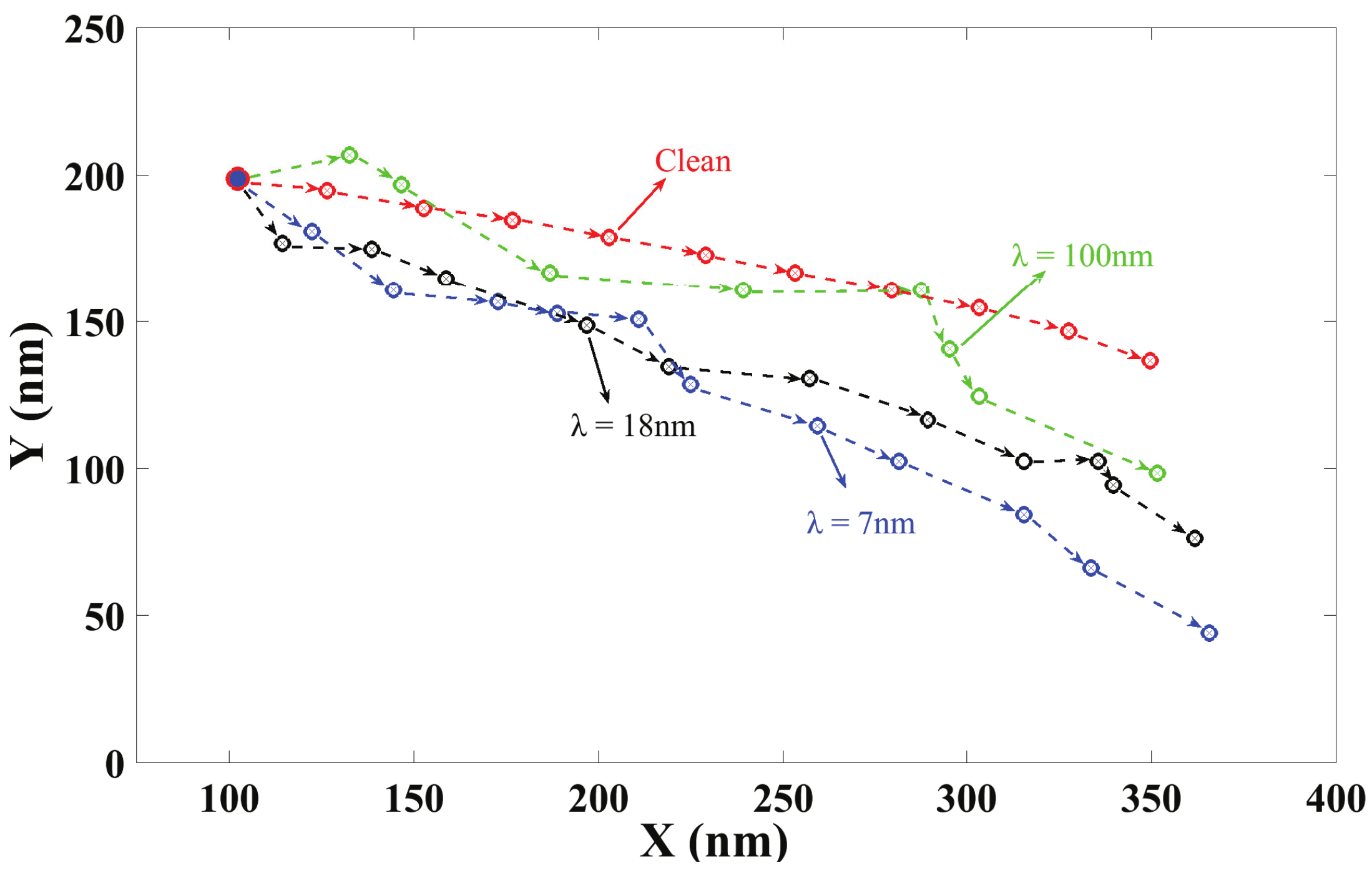}
\caption{(Color online) Simulated skyrmion trajectory for different grain sizes. The non-adiabatic parameter $\beta=0.3<\alpha$. The other parameters are in the main text. The effective skyrmion Hall effect increases with disorder.}
\label{fig:motion}
\end{center}
\end{figure}

The skyrmion is driven by an in-plane spin polarized current along the ${\bf x}$-axis. The skyrmion trajectory for {\color{red}a specific disorder configuration} and different grain sizes near the depinning threshold is shown in Fig. \ref{fig:motion}. The corresponding animations of the skyrmion motion can be found in the Supplementary materials \cite{SuppMat}. In the clean regime, the skyrmion Hall effect is given by $\vartheta_{\rm H}=v_y/v_x\sim\alpha-\beta$ \cite{He2006,37} (see discussion below). However, in the presence of polycrystallinity the skyrmion Hall effect is enhanced, an effect that is prominent in the case of small crystallites and reduced in the case of large crystallites.

\begin{figure}[b]
\centering
\includegraphics[width=0.50\textwidth]{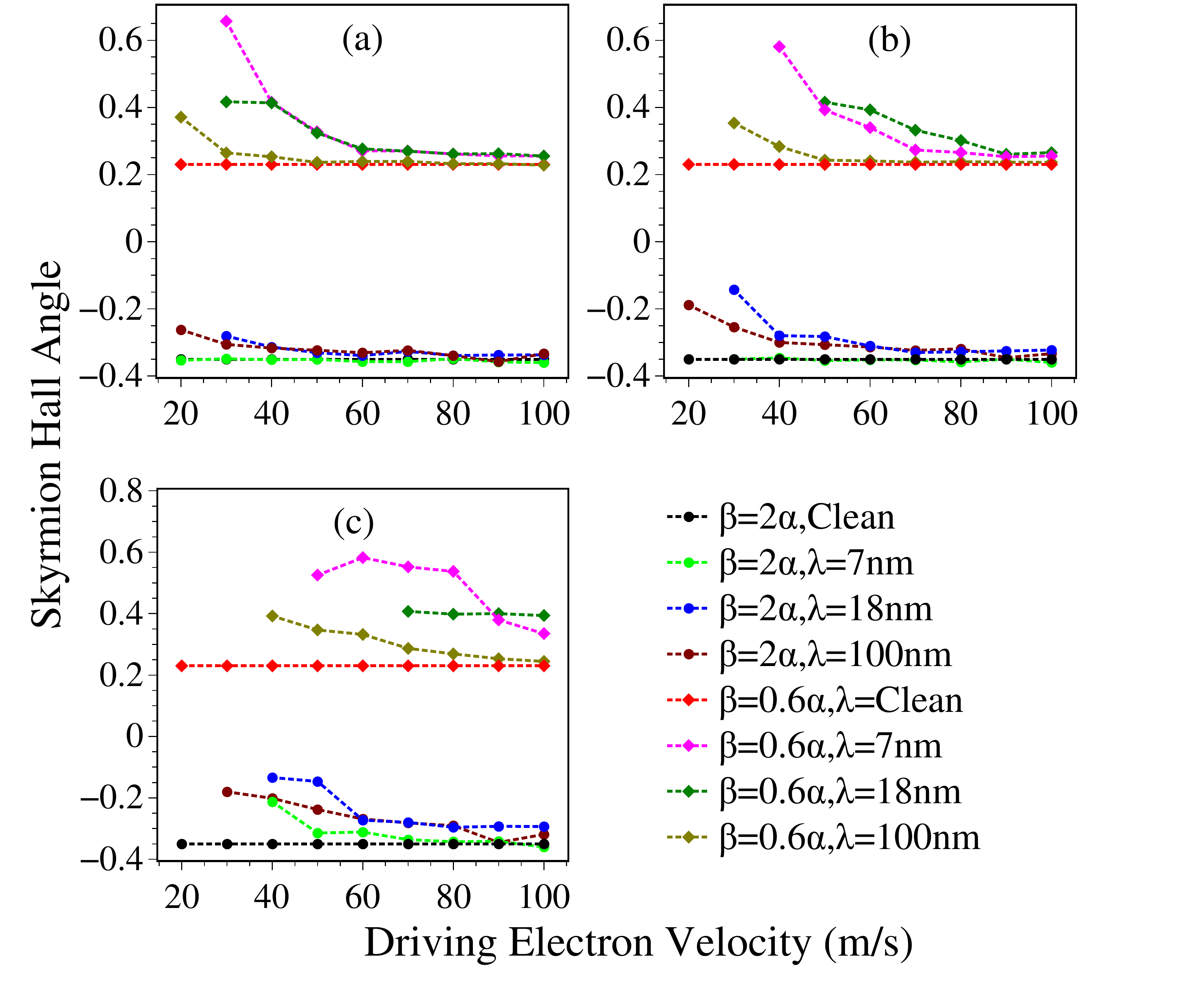}
\caption{Simulated skyrmion Hall angle as a function of current density for different disorder strength for several grain sizes, non-adiabaticity parameter and disorder strength. (a) $\theta$=$2^{0}$, (b) $\theta$= $3^{0}$, (c) $\theta$=$5^{0}$.}%
\label{Fig:Hall_Angle}
\end{figure}

A detailed study of the skyrmion Hall angle as a function of the driving electron velocity $v_s$ is shown in Fig. \ref{Fig:Hall_Angle} for various disorder strengths [(a) $\theta$=$2^{0}$, (b) $\theta$= $3^{0}$, (c) $\theta$=$5^{0}$], and grain sizes ($\lambda$=7, 18, 100 and $+\infty$ nm). For completeness, the simulations were performed for two different non-adiabaticity parameters, $\beta=0.6\alpha$ and $\beta=2\alpha$, corresponding to positive and negative skyrmion Hall effect, respectively. In a clean system, the skyrmion Hall angle is independent of the drive currents for given $\alpha$ and $\beta$ parameters \cite {37}. However, crystallites enhance the skyrmion Hall angle substantially, particularly close to the current threshold. Furthermore, the smaller the grains, the larger the skyrmion Hall angle, in agreement with Fig. \ref{fig:motion}. Interestingly, the skyrmion Hall angle is significantly reduced upon increasing the driving current density, such that in the limit of large driving currents, the skyrmion Hall effect converges towards the results obtained in the clean limit, irrespective of the strength of disorder or the grain size.\par
\begin{figure}[ht]
\begin{center}
\includegraphics[width=0.48\textwidth]{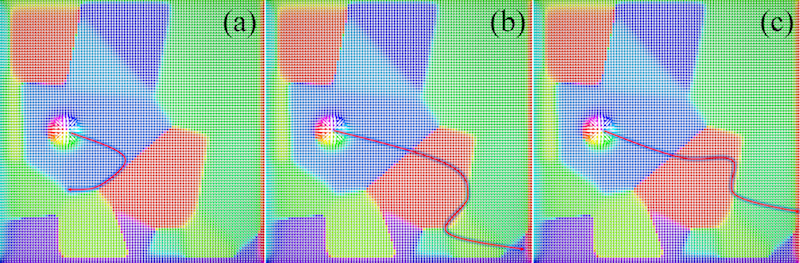}
\includegraphics[width=0.48\textwidth]{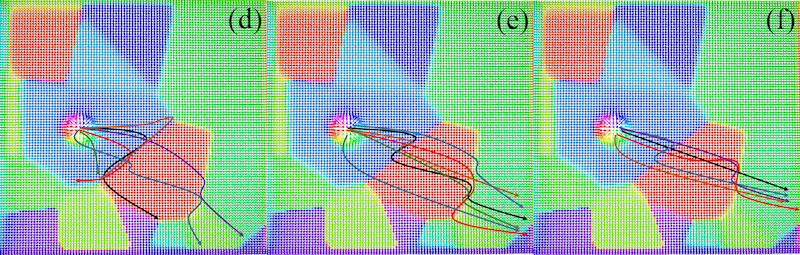}
\includegraphics[width=0.48\textwidth]{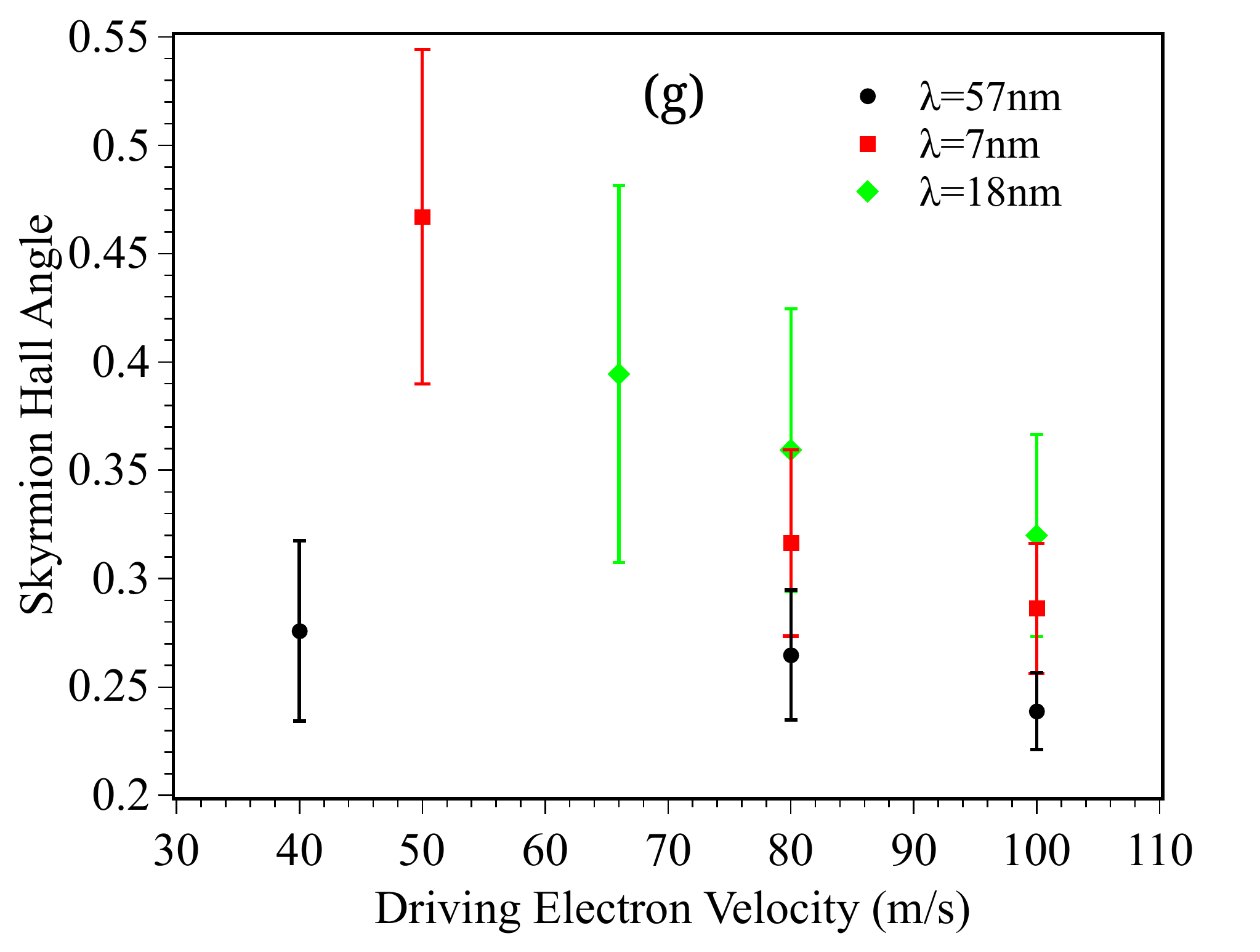}
\caption{(Color online) Skyrmion motion in a polycrystalline sample. (a-c) Skyrmion trajectory in a sample with average grain size 57 nm, disorder strength $\theta$=$5^{0}$ and for increasing driving current densities.(d-f) skyrmion trajectory in a sample with average grain size 57 nm, disorder strength $\theta$=$5^{0}$ and several realizations of anisotropy axis direction for increasing driving current densities.(g) skyrmion Hall angle as a function of current density for disorder strength $\theta$=$5^{0}$ and for several grain size and several realizations of anisotropy axis direction.}
\label{fig:traj}
\end{center}
\end{figure}

To better understand the enhanced skyrmion Hall effect, we investigate the skyrmion trajectory in a polycrystalline landscape. Fig. \ref{fig:traj}(a,b,c) displays the simulated skyrmion trajectory in a sample with an average grain size of 57 nm for different driving current densities. The animations of the simulations are available in the Supplementary materials \cite{SuppMat}. At low current densities [Fig. \ref{fig:traj}(a)], the driving force is insufficient to overcome the repulsive potential formed by the grain boundaries. Consequently, the skyrmion exhibits a sliding motion along the boundary. This motion channels the skyrmion towards the grain junction where it gets pinned. For intermediate current density [Fig. \ref{fig:traj}(b)], the skyrmion passes the first grain boundary and slides on the second grain boundary before getting depinned. By further increasing the current density [Fig. \ref{fig:traj}(c)], the skyrmion becomes less sensitive to the grain boundaries and therefore exhibits a reduced sliding motion. As a result, the overall trajectory of the skyrmion becomes more straight upon increasing the current density, thereby reducing the overall skyrmion Hall effect.\par

{\color{red} In order to test the reproducibility of this behavior, we performed additional simulations on different randomness realizations for a given grain configuration. The results are reported on Fig. \ref{fig:traj}(d,e,f) for six different anisotropy disorder configurations. The other parameters (grain configuration, disorder strength and current densities) are the same as in Fig. \ref{fig:traj}(a,b,c). The skyrmion behavior near the depinning threshold is determined by the grain boundary orientation, resulting in various trajectories depending on the specific disorder configuration. Nevertheless, the basic concept of sliding motion along the grain boundary followed by pinning/depinning at the junctions remains valid. At high current densities though, the trajectory is much less sensitive to the specific disorder configuration and is determined by the driving current density and the non-adiabicity parameter. In Fig. \ref{fig:traj}(g), we show the Hall angle as a function of the the applied current density, averaged over 25 different anisotropy disorder configurations and for various grain sizes. As expected, the Hall angle decreases upon increasing the depinning current density, confirming the effect obtained in Fig. \ref{Fig:Hall_Angle}.}\par

The sliding motion can be accounted for by using Thiele's equation for rigid motion \cite{Thiele1973,He2006,37},
\begin{equation}\label{eq3}
 {\bf G}\times({\bf v}_{s}-{\bf v}_{d})+{\cal D}(\beta{\bf v}_{s} -\alpha{\bf v}_{d})+{\bf F}_{pin}=0,
\end{equation}
where ${\bf G}= {\cal G} {\bf z}$ is the gyromagnetic coupling vector (${\cal G}=4\pi {\cal Q}$, ${\cal Q}$ being the skyrmion charge), ${\cal D}$ is the dissipative tensor (reduced to a scalar by symmetry) and $\mathbf{v}_{d}$ is the drift velocity of the skyrmion. Here ${\bf F}_{pin}=-{\bm\nabla}V_{pin}$ is the pinning force emerging between two adjacent grains, $V_{pin}$ is the pinning potential at the grain boundary. The pinning force is normal to the boundary and can depend on both space and the skyrmion velocity \cite{37}. Rearranging the terms we obtain an expression for the skyrmion drift velocity,
\begin{multline}\label{eq4}
{\bf v}_{d}=\frac{{\cal G}^2+\alpha\beta{\cal D}^2}{{\cal G}^2+\alpha^2{\cal D}^2}{\bf v}_{s} + (\alpha-\beta)\frac{{\cal D}{\cal G}}{{\cal G}^2+\alpha^2{\cal D}^2}({\bf z}\times{{\bf v}_{s}})\\ +\alpha\frac{{\cal D}}{{\cal G}^2+\alpha^2{\cal D}^2}{\bf F}_{pin}-\frac{{\cal G}}{{\cal G}^2+\alpha^2{\cal D}^2}{\bf z}\times{\bf F}_{pin}.
\end{multline}
The first and the second term in Eq. \eqref{eq4} account for the skyrmion longitudinal and transverse velocity components in the disorder free regime, respectively. The third term is the contribution of the repulsive force from the pinning centre and the fourth term corresponds to the gyrotropic motion normal to the direction of pinning force. In the absence of a pinning force, the skyrmion Hall angle is $\vartheta_{\rm H}=v^y_{d}/v^x_{d}=(\alpha-\beta){\cal DG}/({\cal G}^2+\alpha^2{\cal D}^2)$, as mentioned above. However, in the presence of disorder, since the pinning force ${\bf F}_{pin}$ depends on spatial coordinates \cite{32} (and possibly on the drift velocity ${\bf v}_d$ \cite{37}), Eq. \eqref{eq4} is not a closed-form solution and needs to be solved self-consistently. \par

\begin{figure}[ht]
\begin{center}
\includegraphics[width=0.3\textwidth]{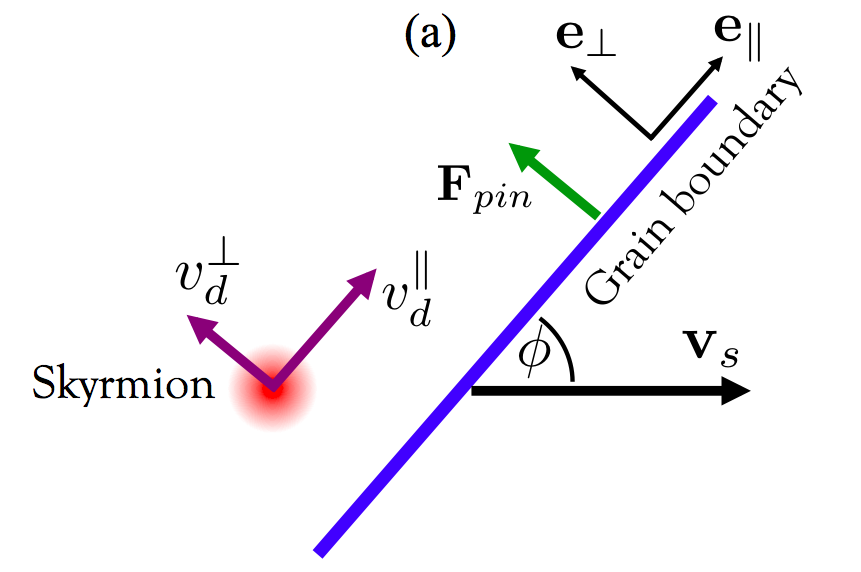}
\includegraphics[width=0.48\textwidth]{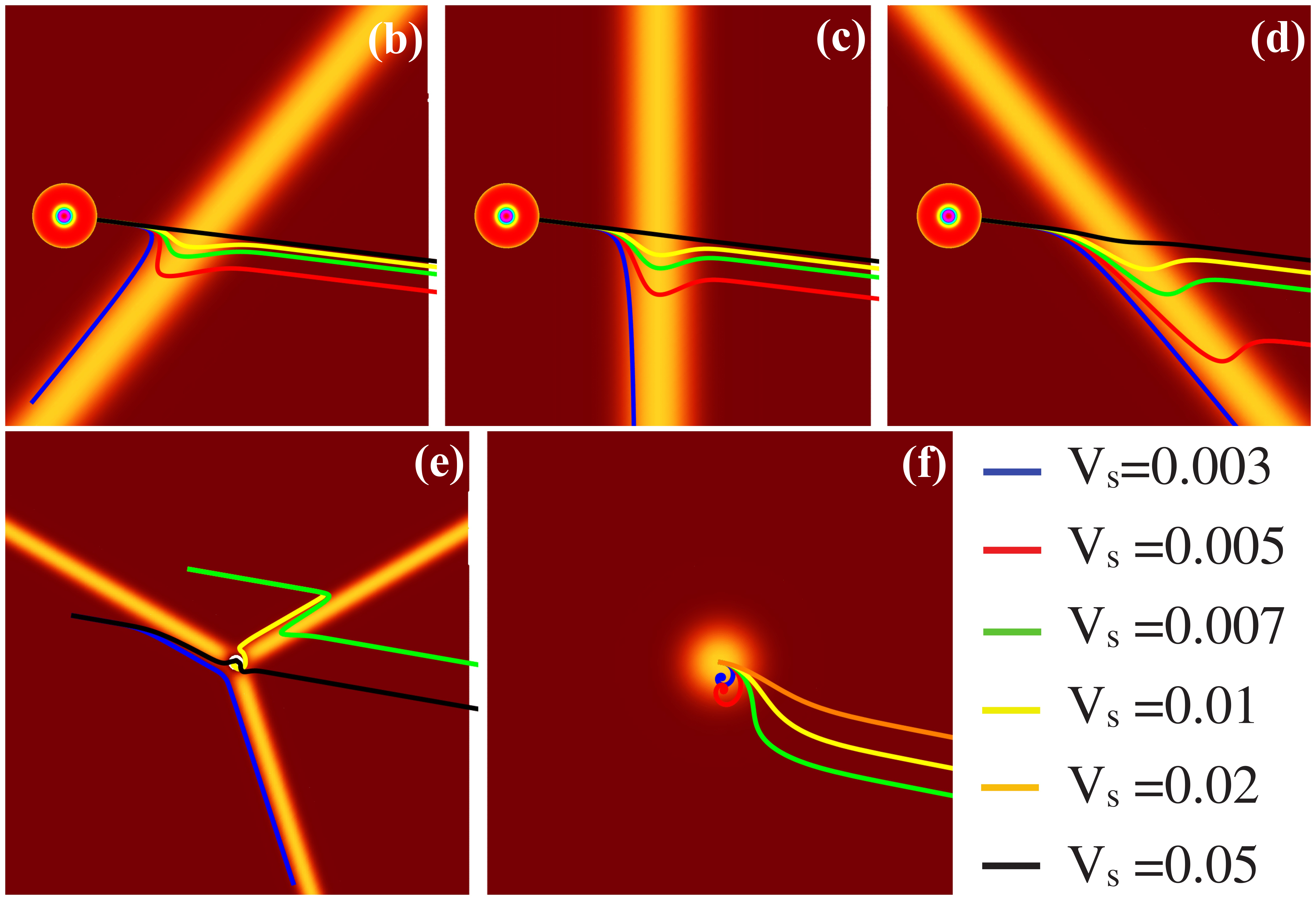}
\caption{(Color online) Skyrmion trajectory along a grain boundary obtained from Thiele's equation. (a) Schematics of skyrmion motion along a grain boundary. (b-e) Skyrmion trajectories obtained by numerically solving Eq. \eqref{eq4} with a Gaussian pinning potential. These calculations describe the skyrmion sliding behavior along different grain boundary orientations. (f) Skyrmion trajectory upon depinning from local defect. We have considered $\beta=0.3$ and assumed ${\cal G}$=${\cal D}$.}
\label{fig:traj2}
\end{center}
\end{figure}

To illustrate the influence of the pinning force on the skyrmion motion, let us consider the case depicted on Fig. \ref{fig:traj2}(a), where a magnetic skyrmion driven by a current along the direction ${\bf v}_s$ impinges on a grain boundary that forms an angle of $\phi$ with ${\bf v}_s$. We parse the skyrmion drift velocity into
\begin{equation}\label{eq5}
{\bf v}_d=v_{d}^\|{\bf e}_\|+v_{d}^\bot{\bf e}_\bot,
\end{equation}
where ${\bf e}_{\|(\bot)}$ is a unit vector along (normal to) the boundary, as indicated on Fig. \ref{fig:traj2}(a). In addition, the pinning force is ${\bf F}_{pin}=F_{pin}({\bf v}_d){\bf e}_\|$. At low current densities, we can impose $v_d^\bot=0$ and deduce the sliding velocity along the boundary,
\begin{equation}\label{eq5}
v_{d}^\|=\frac{{\cal G}\sin\phi+\beta{\cal D}\cos\phi}{\alpha{\cal D}}v_{s}.
\end{equation}
The condition for which the skyrmion crosses the boundary reads $v_d^\bot\leq0$. Therefore, the critical driving current density needed to expel the skyrmion from the grain boundary is
\begin{equation}\label{eq6}
v_s^c=\frac{\alpha{\cal D}}{({\cal G}^2+\alpha\beta{\cal D}^2)\sin\phi-(\alpha-\beta){\cal DG}\cos\phi}F_{pin}^c,
\end{equation}
where $F_{pin}^c=F_{pin}({\bf v}_d^c)$ is the pinning force at the depinning threshold. Hence, the critical skyrmion sliding velocity before it is expelled from the repulsive potential reads
\begin{equation}\label{eq7}
v_d^{\|,c}=\frac{{\cal G}\sin\phi+\beta{\cal D}\cos\phi}{({\cal G}^2+\alpha\beta{\cal D}^2)\sin\phi-(\alpha-\beta){\cal DG}\cos\phi}F_{pin}^c.
\end{equation}
These phenomenological expressions reveal two important features: both the sliding velocity, Eq. \eqref{eq5}, and the critical depinning current, Eq. \eqref{eq6}, depend on the orientation of the grain boundary thereby contributing to a "side jump" of the skyrmion.\par

These effects are further revealed in Fig. \ref{fig:traj2}(b,c,d) which display the skyrmion velocity when approaching a grain boundary for various boundary orientations. These results are obtained by numerically solving Eq. \eqref{eq4} while assuming a quasi 1D repulsive Gaussian pinning potential of the form $V_{pin}(x,y)=V_{0}e^{-\frac{(x\sin\phi-y\cos\phi)^{2}}{R^{2}}}$, where $\phi$ is the angle of the boundary. Here $V_{0}$ is the strength of the interaction and $R$ determines the interaction range. In agreement with our micromagnetic simulations and with the analytical expressions provided above, at low driving current (blue curve) the driving force cannot overcome the potential of the grain boundary, such that the skyrmion remains "stuck" along the boundary, exhibiting the sliding motion discussed above. Upon increasing the driving current, the skyrmion can escape from the grain boundary after a short sliding regime, thereby reducing the effective Hall angle. \par

Interestingly, the distance over which the sliding occurs depends on the angle between the driving current direction and the boundary. This distance is larger for $\phi<0$ [Fig. \ref{fig:traj2}(d)] than for $\phi>0$ [Fig. \ref{fig:traj2}(b)]. In the former case the skyrmion slides along the direction of the driving current, while in the latter case it slides against it. This skewness is directly attributed to the influence of the gyrotropic force $\sim{\cal G}$. Notice also that the sliding distance reduces upon increasing the driving current amplitude, as observed in the micromagnetic simulations in Fig. \ref{fig:traj}.\par

\subsection{Skyrmion depinning}
 
In the previous section, we highlighted that skyrmion sliding along grain boundaries has the tendency of channeling the skyrmion towards grain junctions, as illustrated in Fig. \ref{fig:traj2}(e). Close to such junctions, the skyrmion has a high probability of getting pinned, a probability that decreases upon increasing the driving current density. Let us now investigate the skyrmion depinning behavior (at zero temperature) in polycrystalline samples.\par

\begin{figure}[t]
\begin{center}
\includegraphics[width=0.48\textwidth]{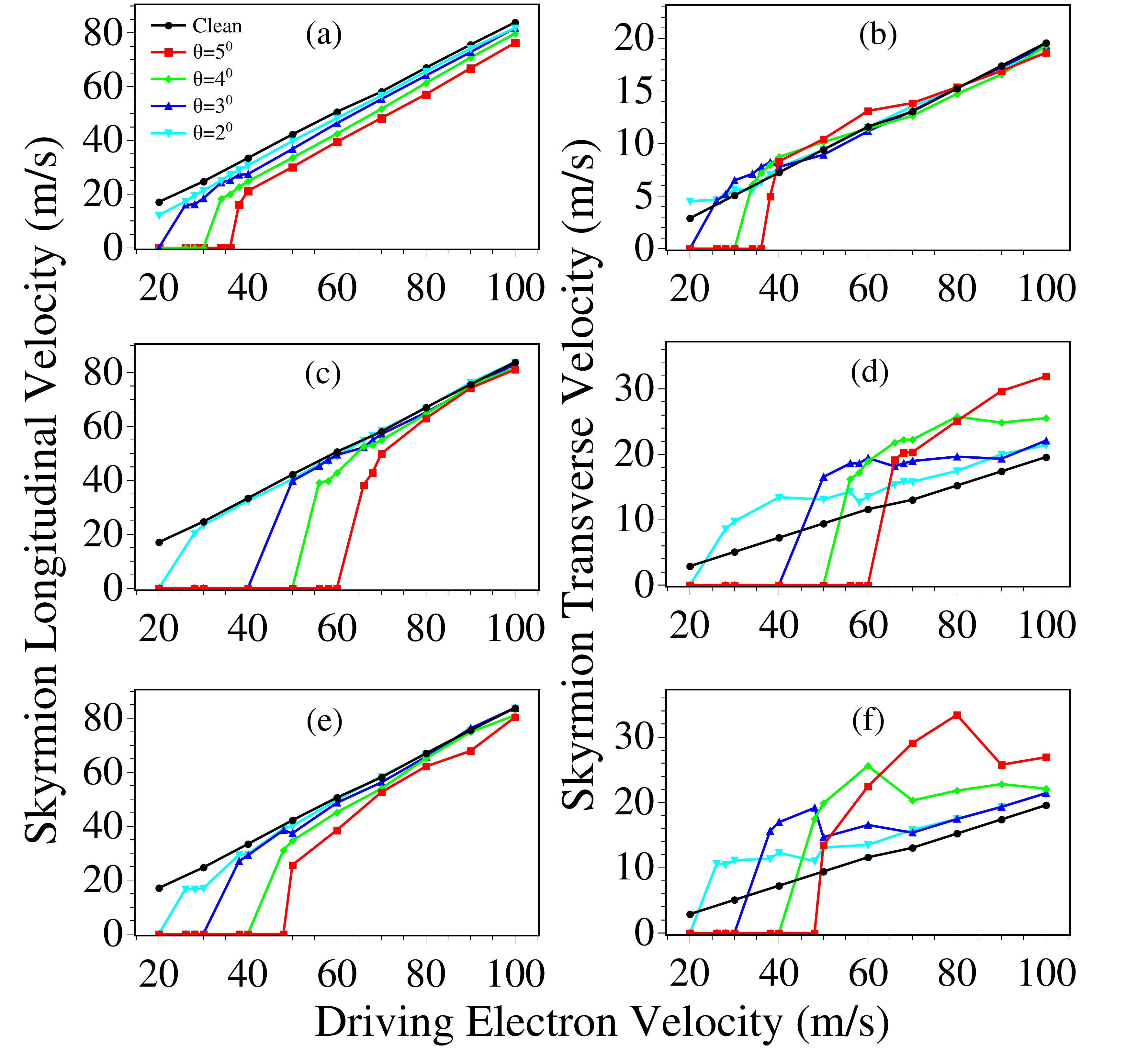}
\caption{Skyrmion longitudinal and transverse velocity components as a function of the driving current for an average grain size of (a,b) $\lambda=100 \text{ nm}$, (c,d) $\lambda=18 \text{ nm}$, and (e,f) $\lambda=7 \text{ nm}$. Here $\beta=0.3$ and the other parameters are given in the main text.}
\label{fig:current}
\end{center}
\end{figure}

\begin{figure}[!t]
\centering
\includegraphics[width=0.23\textwidth]{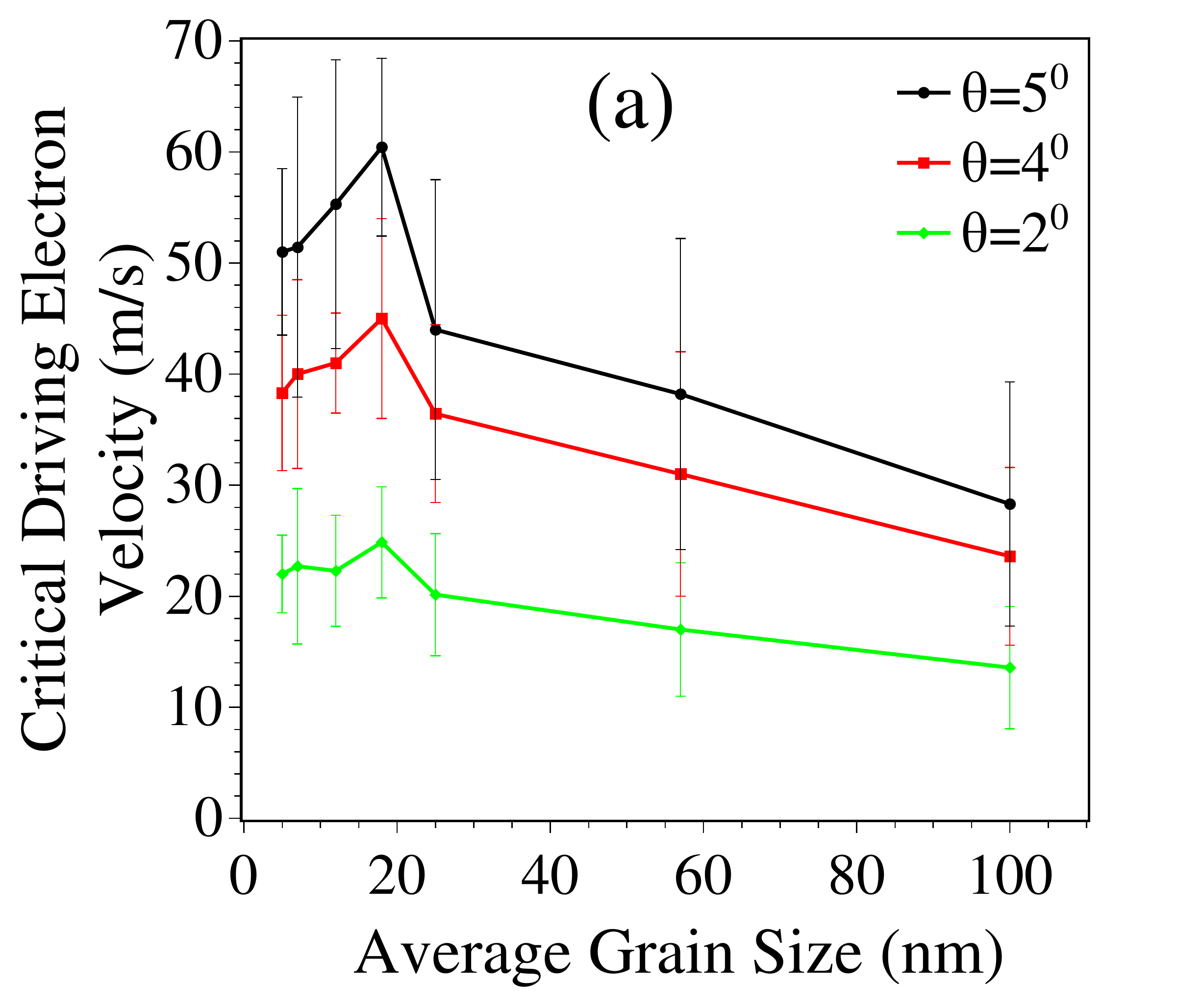}\hspace{2mm}
\includegraphics[width=0.23\textwidth]{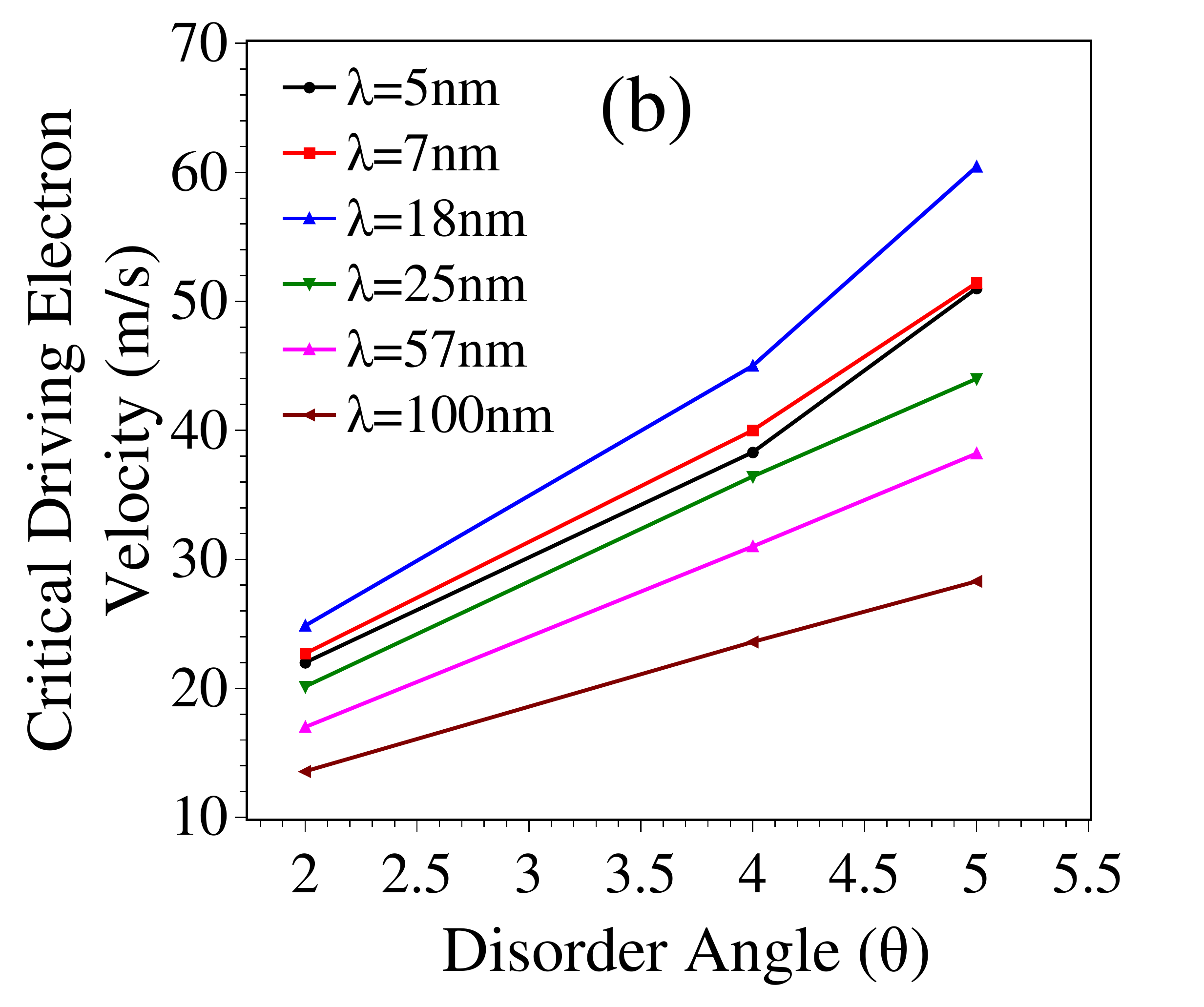}\hspace{2mm}
\caption{(a) Critical current density as function of average grain size. (b) Critical current density as a function of disorder strength defined by $\theta$. Here $\beta=0.3$ and the other parameters are given in the main text.}%.
\label{Velocity}
\end{figure}

Figure \ref{fig:current}(a-f) displays the simulated longitudinal and transverse skyrmion velocity components as a function of the driving current density for the different grain sizes and the disorder strengths. The simulations are performed for the case when $\alpha$ $>$ $\beta$. Depending on the grain size and the disorder strength, we observed significant pinning in these samples. The disorder suppresses the longitudinal velocity component and gives rise to a threshold driving current density, the value of which depends on the grain size, the defect density and the disorder strength [see Fig. \ref{Velocity}(a,b)]. The critical driving current reaches a maximum when the grain size is set comparable to the skyrmion size, as displayed in Fig. \ref{Velocity}(a). This is because the skyrmion then feels the complete impact of each grain. When the grain size is either reduced or enhanced, the critical driving current decreases, an effect also reported in Ref. \onlinecite{26}. In addition, we also observe an enhancement of the skyrmion Hall effect right after the skyrmion gets depinned [Figure \ref{fig:current}(b,d,f)]. This "overshoot" increases upon reducing the grain size and increasing the disorder strength, and decreases upon increasing the driving current density.\par

By analyzing the skyrmion sliding distance and velocity using Thiele's rigid motion model, we can explain the skyrmion trajectory when the average grain size is much larger than the skyrmion size. Since the density of grain junctions is low in this case, the skyrmion speeds up along the grain boundary and most likely leaves it before it reaches the junction, as illustrated in Fig. \ref{fig:traj2}(e). Because of this behavior, the skyrmion has a low probability to get pinned, resulting in a near disorder-free trajectory. At low current density though, the driving force is not large enough to overcome the boundary potential and the skyrmion is likely to be channeled towards the grain junction, leading to enhanced pinning probability. 

Notice that the enhanced skyrmion Hall effect also occurs during the depinning process itself. If one considers a skyrmion originally pinned at a grain boundary or at a local defect, the Hall effect substantially depends on the velocity at which the skyrmion escapes the pinning center. This effect is illustrated on Fig. \ref{fig:traj2}(f) with Thiele's model. We consider a localized pinning potential with cylindrical symmetry and compute the trajectory of the rigid skyrmion as a function of the driving current. Above the threshold, an effect similar to the one reported on Fig. \ref{fig:current} is observed: for a driving current close to the threshold, a "side jump" is observed, which is reduced upon increasing the current.\par

\begin{figure}[t]
\begin{center}
\includegraphics[width=0.49\textwidth]{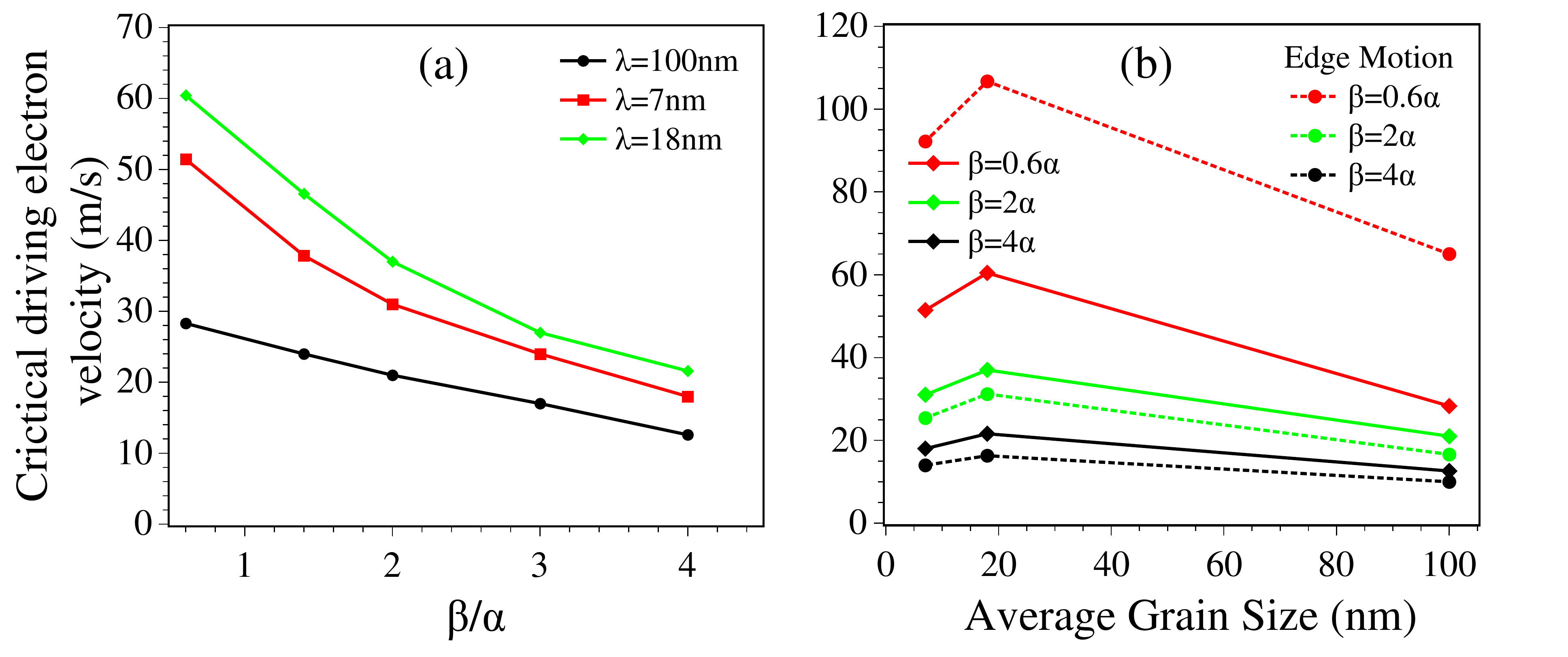}
\caption{(a) Simulated critical current density as a function of non-adiabaticity parameter $\beta$ for various grain size and disorder strength $\theta$=$5^{0}$. (b) Simulated critical current density as a function of average grain size for various $\beta$ parameter. Solid (dashed) lines correspond to a motion taking place near the center (the edge) of the wire.}
\label{fig:beta}
\end{center}
\end{figure}

To complete our discussion, let us turn our attention to the influence of the non-adiabaticity parameter on the critical driving velocity. Indeed, while $\beta$ governs the longitudinal motion of skyrmions only at higher order [$v_d^x=({\cal G}^2+\alpha\beta{\cal D}^2)/({\cal G}^2+\alpha^2{\cal D}^2)$], it drives the transverse velocity when $\beta \neq \alpha$\cite{17,29,Iwasaki2014}. Hence, in a disordered regime dominated by sliding motion along the grain boundaries, increasing $\beta$ drives the skyrmion out of sliding motion and prevents it from being driven into a grain junction resulting in a reduced pinning, as reported in Fig. \ref{fig:beta}(a).\par

The competition between the Magnus force due to the non-adiabaticity parameter and the sliding motion on grain boundaries is best illustrated by considering the skyrmion depinning at the edge of a magnetic wire. There, the Magnus force due to the non-adiabaticity parameter is balanced by the confinement force. Fig. \ref{fig:beta}(b) displays the critical current density as a function of the grain size for skyrmions initially located at the center (solid lines) and at the edges of the sample (dashed lines). When $\alpha$ $<$ $\beta$ (green and black symbols), the skyrmion is pushed towards the edge and slides along it with velocity $v_d^x=(\beta/\alpha) v_s$\cite{28,29,Iwasaki2014}, resulting in a reduced critical driving current density compared to the case of central motion. However, when $\alpha$ $>$ $\beta$ the skyrmion is pushed along the opposite edge of the sample, such that the influence of the confinement force is reduced. As a result, the skyrmion is more sensitive to defects and grain boundaries, leading to higher threshold current density. These observations suggest that narrow nanowires have the potential to enable high skyrmion mobility, not only by quenching the non-adiabatic Magnus force, but also by reducing the sensitivity of skyrmions to pinning.

\subsection{Impact of temperature\label{s:thcom}}

{\color{red} We emphasize that the effects discussed above do not account for the impact of thermal activation. Brownian motion of skyrmions \cite{Troncoso2014,Schutte2014,Rohart2016,Miltat2018,Rozsa2016} and mechanisms for thermally-induced skyrmion annihilation in clean ferromagnets have been investigated by various groups \cite{Rohart2016,Lobanov2016}. In particular, the thermal stability of skyrmions in nanotracks has been scrutinized and it has been shown that the thermal activation energy is substantially reduced at the boundary of the track compared to the one inside the track. As a result, while thermal fluctuations reduce the overall skyrmion stability, it is even more dramatic at boundaries due in part to the canting of the magnetic moments \cite{Stosic2017,Cortez2017,Bessarab2018}. The same applies to non-magnetic impurities \cite{Uzdin2018a,Uzdin2018b}, and therefore, one can speculate that thermal fluctuations reduce the skyrmion stability at the grain boundaries.}
 {\color{red} Temperature effects on skyrmion stability have been reported earlier \cite{Miltat2018}. Thermal diffusion can cause significant skyrmion distortion resulting in reduced stability. Furthermore, the thermal fluctuations can weaken the skyrmion-grain boundary interaction. Therefore, the temperature is expected to favor a low critical current density. \cite{Iwasaki2014}}. 
\section{Conclusion}
We have investigated the role of granular boundaries on the skyrmion motion in polycrystalline samples through micromagnetic simulations and Thiele formalism. We found that the critical current density depends on the relative size between the grains and the skyrmion and reaches a maximum when both are of the same order. More intriguingly, we observe that the skyrmion Hall effect is effectively enhanced by the presence of disorder. This is caused by the onset of sliding motion of the skyrmion along the grain boundaries, and side jump from grain junctions. Such an effect is overall quenched upon increasing the driving current density. Finally, we have shown that when the skyrmion trajectory is confined along an edge by the non-adiabatic Magnus force, the critical current density can be significantly reduced, suggesting that narrow nanowires yield higher skyrmion mobility.\\

A.S., A.A. and A.M. acknowledge support from the King Abdullah University of Science and Technology (KAUST), as well as computing time on the supercomputers SHAHEEN at KAUST Supercomputing Center and the team assistance. A.B. is supported by the Research Council of Norway through its Centre of Excellence funding scheme, Project No. 262633, "QuSpin".

\end{document}